\documentclass[conference]{IEEEtran}
\IEEEoverridecommandlockouts
\usepackage{cite}
\usepackage{amsmath,amssymb,amsfonts}
\usepackage{algorithmic}
\usepackage{graphicx}
\usepackage{textcomp}
\usepackage[table]{xcolor}
\usepackage{hyperref}
\usepackage{balance}

\usepackage{graphicx}
\usepackage{subcaption}
\usepackage{enumitem}
\usepackage{float}
\usepackage{xcolor}
\usepackage[export]{adjustbox}
\usepackage{float}
\usepackage{tabulary}
\usepackage{tabularray}
\usepackage{url}
\usepackage{blindtext}
\usepackage{tcolorbox}
\usepackage{graphicx}
\usepackage{wrapfig}
\usepackage[T1]{fontenc}

\usepackage{listings}
\usepackage{subcaption}
\usepackage{xspace}

\newcommand*{\ie}{i.e.,\@\xspace}
\newcommand*{\eg}{e.g.,\@\xspace}

\newcommand*{\iot}{IoT\@\xspace}
\newcommand*{\RD}{ResyDuo\@\xspace}
\newcommand*{\PH}{ProjectHub\@\xspace}
\newcommand*{\AD}{Arduino\@\xspace}

\newcommand{\rqfirst}{\textbf{RQ$_1$}:\textit{ How effective is \RD's \typeI recommendation component?}~}

\newcommand{\rqsecond}{\textbf{RQ$_2$}:\textit{How effective is \RD's \typeII recommendation component?}~}%How do the recommendation engine improvements impact on the \MGp's performances, including preprocessing and vertex histogram kernel function?
\newcommand{\rqthird}{\textbf{RQ$_3$}:\textit{ How effective is \RD's \typeIII recommendation component?}~}

\newcommand*{\typeI}{Type I\@\xspace}
\newcommand*{\typeII}{Type II\@\xspace}
\newcommand*{\typeIII}{Type III\@\xspace}

\makeatletter
\newcommand*{\etc}{%<
	\@ifnextchar{.}%
	{etc}%
	{etc.\@\xspace}%
}
\makeatother
\newcommand*{\etal}{\textit{et~al.}\@\xspace}

\definecolor{verylightgray}{gray}{0.95}

\newcommand\cready[1]{\textcolor{black}{#1}}
\def\BibTeX{{\rm B\kern-.05em{\sc i\kern-.025em b}\kern-.08em
    T\kern-.1667em\lower.7ex\hbox{E}\kern-.125emX}}
\begin{document}

\makeatletter
\newcommand{\linebreakand}{%
  \end{@IEEEauthorhalign}
  \hfill\mbox{}\par
  \mbox{}\hfill\begin{@IEEEauthorhalign}
}
\makeatother

\title{\RD: A Recommender System for Developing \AD-based projects}

\title{\RD: Combining data models and CF-based recommender systems to develop \AD projects}

%\title{\RD: Enabling CF-based recommender systems with data model}

%\and
%\and
%\and
%\and
%\and

\author{\IEEEauthorblockN{Juri Di Rocco}
\IEEEauthorblockA{\textit{University of l'Aquila} \\
L'Aquila, Italy \\
juri.dirocco@univaq.it}
\and
\IEEEauthorblockN{Claudio Di Sipio}
\IEEEauthorblockA{\textit{University of l'Aquila} \\
L'Aquila, Italy\\
claudio.disipio@univaq.it}
}

\maketitle

\begin{abstract}
 \cready{While specifying an IoT-based system, software developers have to face a set of challenges, spanning from selecting the hardware components to writing the actual source code. Even though dedicated development environments are in place, a non-expert user might struggle with the over-choice problem in selecting the proper component. 
 By combining MDE and recommender systems, this paper proposes an initial prototype, called \RD, to assist \AD developers by providing two different artifacts, \ie hardware components and software libraries. In particular, we make use of a widely adopted collaborative filtering algorithm by collecting relevant information by means of a dedicated data model. \RD can retrieve hardware components by using tags or existing \AD projects stored on the \PH repository. Then, the system can eventually retrieve corresponding software libraries based on the identified hardware devices. \RD is equipped with a web-based interface that allows users to easily select and configure the under-developing \AD project. To assess \RD's performances, we run the ten-fold cross-validation by adopting the grid search strategy to optimize the hyperparameters of the CF-based algorithm. 
 The conducted evaluation shows encouraging results even though there is still room for improvement in terms of the examined metrics.}
\end{abstract}


\section{Introduction}
\label{sec:introduction}
\cready{The Internet of Things (\iot) paradigm involves the communication among different physical objects that collect, process, and exchange real-time data \cite{bandyopadhyay_internet_2011,atzori_internet_2010}. In a broader sense, an \iot application can be defined as a networked infrastructure of heterogeneous embedded systems, including sensors, actuators, and other physical devices \cite{samie_iot_2016}. Their development encompasses the selection of hardware and software components that need to be properly integrated to realize the wanted functionalities. In particular, an activity of paramount importance is the mapping between the chosen hardware components and the corresponding software libraries to avoid unexpected behaviors \cite{9402092}. }

%Such systems have become pervasive in multi-disciplinary domains, \eg intelligent transportation  systems~\cite{7152667}, healthcare~\cite{doi:10.1155/2014/217415}, and smart business~\cite{nieto2009data}. 
Although several modeling techniques have been recently proposed to assist modelers in specifying and developing \iot applications \cite{mijalkovic_traceability_2022,alfonso_model-based_2023,Ihirwe2020},  there is still the need to support developers by considering additional features, \eg supporting the mapping between hardware and software components. In this respect, recommender systems (RSs) are widely adopted to facilitate the development of complex software systems \cite{RSSE_BOOK_2014,ricci_recommender_2011,savary-leblanc_software_2023} by providing different artifacts.

In this paper, we focus on the development of a specific class of embedded systems, namely \AD-based ones, by combining the MDE paradigm and mining techniques to devise \RD, a \textbf{RE}commender \textbf{SY}stem for ar\textbf{DU}in\textbf{O} project built on top of a tailored data model and collaborative filtering (CF) strategy. 
First, we mined the \PH repository \cite{projectHub} to extract relevant metadata related to both users and projects, \ie hardware components, tags, and software libraries. To this end, we use a tailored data model to filter the mined data according to some quality aspects, \eg removing infrequent projects. 
%By relying on the collected data, the CF-based engine can first support the hardware configuration phase by exploiting \textit{i)} the tags of the project or \textit{ii)} the existing list of components. Once those components have been selected, \RD can infer the corresponding software libraries to actually develop the \AD-based project. 
\cready{By relying on the collected data, the CF-based engine can support both hardware 
 and software configuration phases by exploiting \textit{i)} the tags of the project, \textit{ii)} the existing list of hardware components, and \textit{iii)} inferring the corresponding software libraries.
All the recommended items are eventually retrieved to the user using a dedicated web interface equipped with an integrated code editor.  Our approach focuses on representing \iot data related to the \PH repository. In particular, such an abstraction enables the mining activities that are needed to feed the underpinning CF system, \ie three different projections are extracted from \PH raw data to support the three different recommendations.}  
Due to the lack of a proper baseline, we automatically evaluate the system by using the cross-fold validation strategy \cite{RefaeilzadehTL09}. Furthermore, we adopt the well-known grid search algorithm \cite{automatedML_2019} to select the proper hyperparameters of the CF-based engine. Although the results are low in terms of the metrics examined, it is our strong belief that \RD can support by providing effective recommendations.
The contributions of this paper are the following:
%Thus \textit{ReSyDuo} is capable of guiding the user during the development of Arduino prototypes by providing recommendations of the hardware components and third-party libraries to be used. To realize \textit{ReSyDuo} it was necessary to capture not only the technical aspects but also aspects related to the Arduino community, with information pertaining to user behavior regarding categorization projects with user-defined tags \cite{zuo2016tag}. The techniques used in the e-commerce's recommender system have been adapted for the specific Arduino domain. Finally some approaches used in the \textit{Recommender Systems in Software Engineering} have been useful for the recommendation of libraries, as seen in CrossRec \cite{CrossRec_2021}.\\

\begin{itemize}
\item[--]  A dedicated metamodel that enables mining and filtering operations on extracted Arduino projects;
\item[--]  A recommender system called \RD that can assist \AD developers in specifying the mapping between hardware components and external software libraries by using three different types of recommendations;
\item[--]  A quantitative evaluation exploiting well-founded metrics and a grid-search strategy to optimize the algorithm parameters 
\item[--] \cready{A replication package to foster further research in the domain \footnote{\url{https://github.com/MDEGroup/low-code-ard-proj}}}
\end{itemize}

This paper is structured as follows. Section \ref{sec:background} presents a motivation example and the technical background of the paper. The \RD architecture and its subcomponents are discussed in Section \ref{sec:approach}. Section \ref{sec:evaluation} and Section \ref{sec:results} describe the evaluation process and the obtained results respectively. Threats that may compromise the effectiveness of the system have been discussed in Section \ref{sec:threats}. We eventually conclude the paper in Section \ref{sec:conclusion} by summarizing the contributions and envisioning possible future works. 

%background, provides an overview of a state of the art for the Recommender Systems, shows the techniques used, and clarifies some basic concepts present in the thesis. In Chapter \ref{ReSyDuo}, the proposed approach is shown. Chapter \ref{Evaluation} presents the evaluation methodology used to validate the ReSyDuo.  The Chapter \ref{Result} show the result obtained by ReSyDuo in the validation task. Chapter \ref{RelatedWork} contains related work used in this thesis. Finally, Chapter \ref{Conclusions} concludes the thesis.

\section{Background and Motivating example}
\label{sec:background}

This section is two-fold. First, we introduce the Arduino programming and the \PH ecosystem in Section \ref{sec:ecosystem}. 
Section~\ref{sec:mde-iot} presents of the MDE approaches can support the development of \iot systems.
A brief overview of collaborative filtering techniques is provided in Section~\ref{sec:collaborative-filtering}.
Finally, Section \ref{sec:motivating} presents a motivating example to discuss the peculiarities and challenges of developing IoT devices. 
%We eventually describe the collaborative filtering strategy employed in Section \ref{sec:cf}. 

\subsection{The Arduino Ecosystem}\label{sec:ecosystem}

Among the others, the Arduino hardware platform is the most adopted by the IoT community since \textit{i)} it requires programming skills or domain-specific knowledge and \textit{ii)} it provides several open-source libraries and projects \cite{noauthor_arduino_nodate}.
%Arduino’s little, blue circuit board, mythically taking its name from a local pub in Italy,  it has motivated a new generation of DIYers of all ages to make all manner of wild projects found anywhere from the hallowed grounds of our universities to the scorching desert sands of a particularly infamous yearly arts festival and just about everywhere in between. Usually, these Arduino projects require little programming skills or knowledge of electronics theory, and more often, this handiness is picked up along the way.
%Arduino is not a simple development environment like Java or any other programming language that allows you to develop open-source code and share it. 
%Arduino introduces the concept of open-source hardware. In the past, we used to hear about open-source software but never about open-source hardware. In agreement with Evans, in the book Beginning Arduino Programming \cite{BeginningArduinoProgramming}.
The development process involves two main types of layers, \ie the \textit{physical} and the \textit{software} one. The former requires the configuration of hardware components, \eg sensors or actuators. The latter manages the interaction among those elements by using a derived language of C++. 
%onsisting of a microcontroller where analogs and digital hardware components can be plugged, i.e., ; the \textit{software layer} uses a derived language of C++. Libraries allow for reusing code to manage hardware components.
%Figure \ref{fig:ArduinoIDEHelloWorld} show a simple \textit{Hello World} program for Arduino written  with the Arduino IDE.
%The code of an Arduino program has a well-defined structure; it can be divided into three sections: At the top, there is the \textit{header} section where the user can include libraries and define variables. The second section is the \texttt{setup()} function,  which is used to initialize variables and verify the correct functioning of the hardware components. Finally, in the third section, inside the \texttt{loop()} function, we can define the behavior of our program. This function is performed continuously, so it will also be necessary to manage the pauses between one action and another, for example, between reading a sensor and executing the `action' of an actuator.

A crucial part of coding with Arduino is the ability to reuse code, which can be accomplished through the use of third-party libraries (TPLs). These libraries provide commonly used features that implement the behavior of a single \iot component.

%The simplicity of development in Arduino and its open-source nature has brought many people closer to this ecosystem; the community has grown and contributed to developing new technologies compatible with Arduino, expanding the software and hardware panorama.

%\subsection{Repositories for Arduino Projects} 
%The first phase of the analysis consisted in %finding a data source; the requirement that %this must satisfy is to provide a sufficient %number of samples. There are many online %repositories, such as GitHub, ProjectHub, %GitLab, and others.
%Summarizing, the most important characteristics %that the repository must satisfy are the %following:
%\begin{itemize}[noitemsep]
%    \item[--]semantic information;
%    \item[--]hardware information;
%    \item[--]software information.
%\end{itemize}

%Similar to the approach proposed by Di Ruscio %et al. in \cite{Di_Ruscio_2014} an analysis has %been performed on various repositories that %could satisfy the characteristics listed above.
%In the solution proposed by Di Ruscio et al, %several repositories have been analyzed, and a %meta-model has been created which allows %modeling the information of the repositories.
%The border ucommunity of arduino developers shares their projects on 

To assist \AD developers in selecting the proper components,  the \PH \cite{projectHub} repository collects, stores, and maintains Arduino projects developed by the community, thus easing the burden of choice. The \textit{Project Header} section contains all the relevant metadata, including title, description, tags, views, and comments. Users can search the list of hardware components used in the project by navigating the \textit{Components and Supplies} section. %For some projects, external \textit{Tools and Machines} are required to build the prototype, \eg the scissors.  and finally the \textit{Source Code} of the project. 
Each project has the corresponding \textit{source code} that can be downloaded and integrated using dedicated development environments, like Arduino IDE.

Furthermore, projects are typically classified by \textit{tags}, a list of keywords that represent the content of repositories, \eg \textit{Smart Agriculture}, \textit{Health},  or \textit{Temperature Measurements}. Similar to GitHub topics\footnote{\url{https://github.com/topics}}, such keywords can improve the discoverability of a particular project and help users in finding similar projects for their needs.

Despite the \AD project being well-structured, a newcomer developer may waste time gathering relevant information directly from \PH. In this paper, we conceive a dedicated data model that can be used to represent and mine the needed data easily.

%A user can find useful information in multiple projects, such as information about displaying data in one project and information about using a sensor to detect the temperature in another. 

\subsection{Applyting MDE to engineer \iot applications}\label{sec:mde-iot}

MDE has been applied to CPSs to facilitate their design and development. ThingML \cite{ThingMLcore} is an \iot platform that exploits imperative actions to develop \iot projects. The platform exploits UML models to represent the components and their behavior. A dedicated DSL language has been conceived to support the specification of wastewater treatment plant (WWTP) systems \cite{costa2016modeling}. Furthermore, the framework is capable of generating the corresponding source code, thus supporting the specification of the system both at design and run time. Since \iot systems are composed of heterogeneous components, several approaches focus on specific aspects, \ie specification of critical missions \cite{Ciccozzi2017}, traceability \cite{mijalkovic_traceability_2022}, or testing \cite{cristea_building_2022}.
Besides traditional techniques, multi-modeling paradigm (MPM) \cite{amrani_multi-paradigm_2021}, digital twins \cite{EramoBCBWW22}, and DevOps \cite{colantoni2020devopsml} can be exploited to automatize the design of CPS by combining different MDE artifacts.  

\subsection{Collaborative filtering}\label{sec:collaborative-filtering}
Collaborative filtering is a well-known technique in the field of recommender systems that has revolutionized the way we obtain individualized recommendations across a variety of online platforms~\cite{Schafer:2007:CFR:1768197.1768208}. This clever technique uses collective intelligence from a large pool of users to give personalized recommendations to individuals.

To create forecasts and suggestions, collaborative filtering depends on the wisdom of the crowd. It evaluates the behavior and preferences of many users to uncover patterns and similarities, rather than relying simply on explicit item features or user preferences~\cite{di2015recommender}. %Collaborative filtering algorithms can easily connect like-minded individuals and offer goods that coincide with their interests by effectively mining the collective knowledge of these users.
The two most common ways of collaborative filtering are user-based~\cite{zhao2010user} and item-based~\cite{sarwar2001item}. The user-based method discovers users with similar preferences and proposes things that these users have loved. The item-based method, on the other hand, focuses on discovering commonalities between things based on the people who have shown a liking for them. Both strategies use user-item interaction data to form meaningful relationships, resulting in accurate and appropriate suggestions~\cite{Schafer:2007:CFR:1768197.1768208}. To the best of our knowledge, none of the existing approaches cover the representation of \AD data. Although the application domain is limited, the proposed data model can be enriched by mining additional data sources.

\subsection{Motivating Example}\label{sec:motivating}
%The Arduino team realized that people don’t care about the inner workings of a microcontroller board but rather want to develop something cool. Therefore, the Arduino team has created a user-friendly environment. Being an open-source environment, the number of community users has grown exponentially over time, leading to a growth in the number of available hardware and software components. A developer must choose both the hardware component and the library, which may be a complex task. Therefore, it is necessary to import a library or manually implement the code for hardware management for each hardware component; for instance, a temperature sensor requires its own library or a compatible one. It is possible to find different hardware versions and libraries that are compatible with them.  Let's describe a practical example of the typical Arduino development scenario. 

%A user must first define the problem he wishes to solve; for example, suppose a user is developing a prototype to monitor a small vegetable farm similar to the prototype presented by Akbar et al in \cite{TempHum2020prototype}. 

\begin{figure*}[h!]
    \centering
    \begin{subfigure}[b]{0.45\linewidth}
        \includegraphics[width=\linewidth]{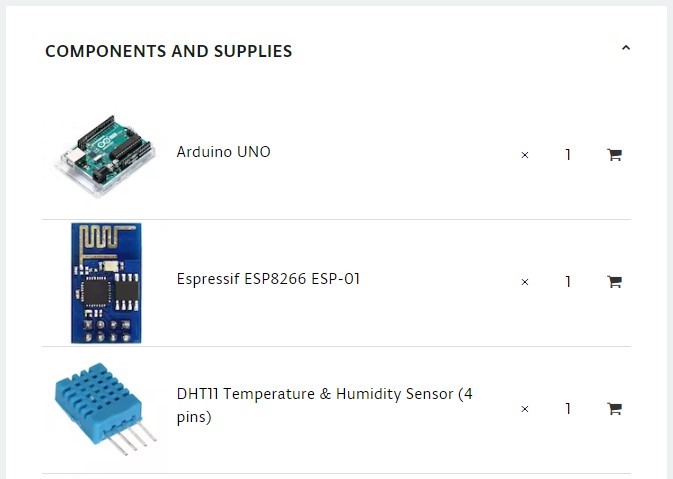}
       
        \caption{Hardware configuration phase.} \label{fig:part1}
    \end{subfigure}
    \hspace{0cm}
    \begin{subfigure}[b]{0.45\linewidth}
        \includegraphics[width=\linewidth]{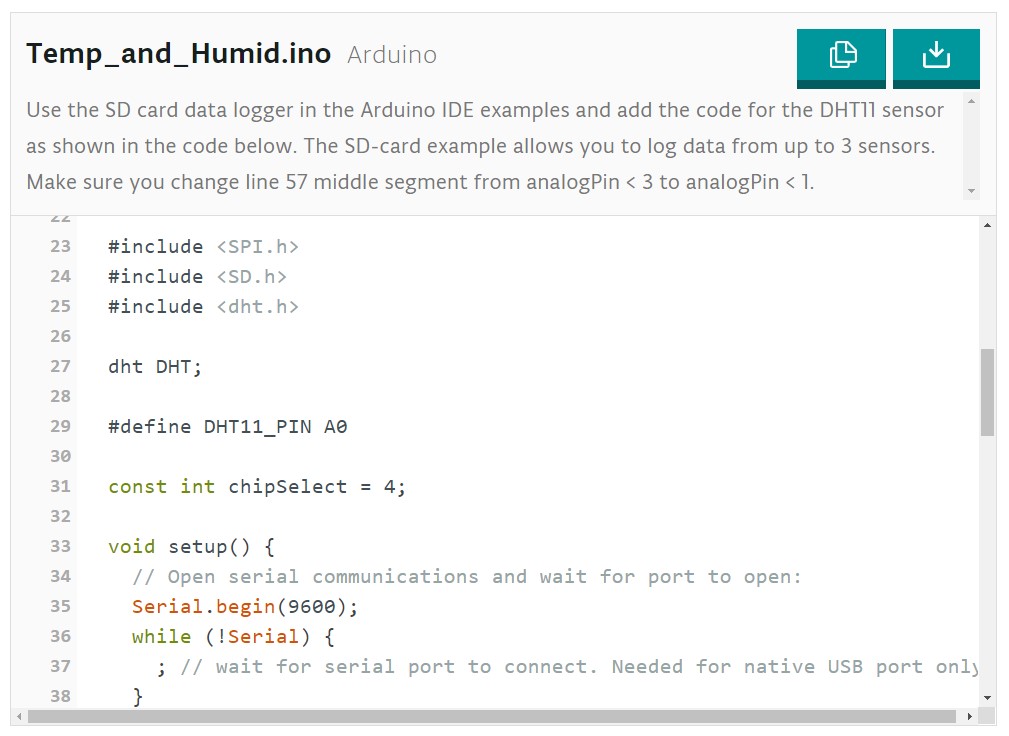}
     
        \caption{Software development.}\label{fig:part2}
    \end{subfigure}
    \caption{Motivating example.}
    \label{fig:motivating}
\end{figure*}

Figure \ref{fig:motivating} depicts the development process by using \AD IDE. As the first step, the user has to select a proper list of hardware components according to the type of \AD board employed as shown in Figure \ref{fig:part1}. This selection is bounded by the application domain and it is crucial to realize the functionalities elicited at the design time. Then, the user can employ a dedicated IDE to develop the actual code. Figure \ref{fig:part2} depicts an explanatory configuration file in which the user has to map the selected hardware components to the corresponding software ones, \ie external libraries. Even though dedicated environments like \AD IDE \footnote{\url{https://www.arduino.cc/en/software}} can facilitate the system's specification, empirical studies demonstrate that users encounter issues during the development phase \cite{10.1145/2858036.2858533}. In particular, a wrong mapping between hardware and software components may lead to unexpected behaviors \cite{9402092}.

To cope with those limitations, we propose a recommender system based on CF algorithms that exploit a tailored metamodel. In the next section, we detail all the phases supported by \RD, from the data collection to the actual recommendations.

\section{Proposed Approach}
\label{sec:approach}

\RD aims at supporting different types of recommendations as follows:
\begin{itemize}

 %this type of recommendation needs the %entire project as input and receives a set of recommended components that are used in similar projects.
   \item[--] \textbf{\typeI:} Hardware components from tags: given an empty project, the user can specify the tag to collect initial recommendations in terms of hardware components

    \item[--] \textbf{\typeII:} Hardware components from projects:  
    %These recommendations can be generated for projects where some components have already been chosen. 
    This type of recommendation aims to enhance the project under development by retrieving components from similar projects.  
   \item[--] \textbf{\typeIII:} Software libraries from components: Once the hardware components have been selected, this type of recommendation can provide the software components to develop the actual system 
\end{itemize}

It is worth mentioning that \typeI and \typeII recommendations aim to recommend hardware components, they are conceived to support two different phases of implementing Arduino projects, \ie early configuration phase and project development phase. Once components have been selected, the \typeIII recommendations suggest suitable libraries to interact with hardware components. %retrieves the corresponding TPLs to complete the actual software implementations.   

Figure \ref{fig:ResyduoOverView} depicts the \RD overall architecture that consists of two different sub-processes, \ie the \textit{Building phase} and \textit{Deployment phase}. In the former, the system extracts relevant data from the \PH repository and encodes the relevant data by relying on \texttt{Data Scraper} and \texttt{Data Miner} components respectively. In the latter, \RD relies on the gathered data to feed the \texttt{Recommender component} that retrieves the recommender items in a dedicated \texttt{Presentation} REST interface.

%First, the \textit{Data miner} module extracts and encodes the relevant data by relying on a tailored data model. Such preprocessing step enables the \textit{Recommender engine} component that assesses the similarity between the active context and the mined knowledge base. 

\begin{figure}
    \centering
    \includegraphics[width=0.5\textwidth]{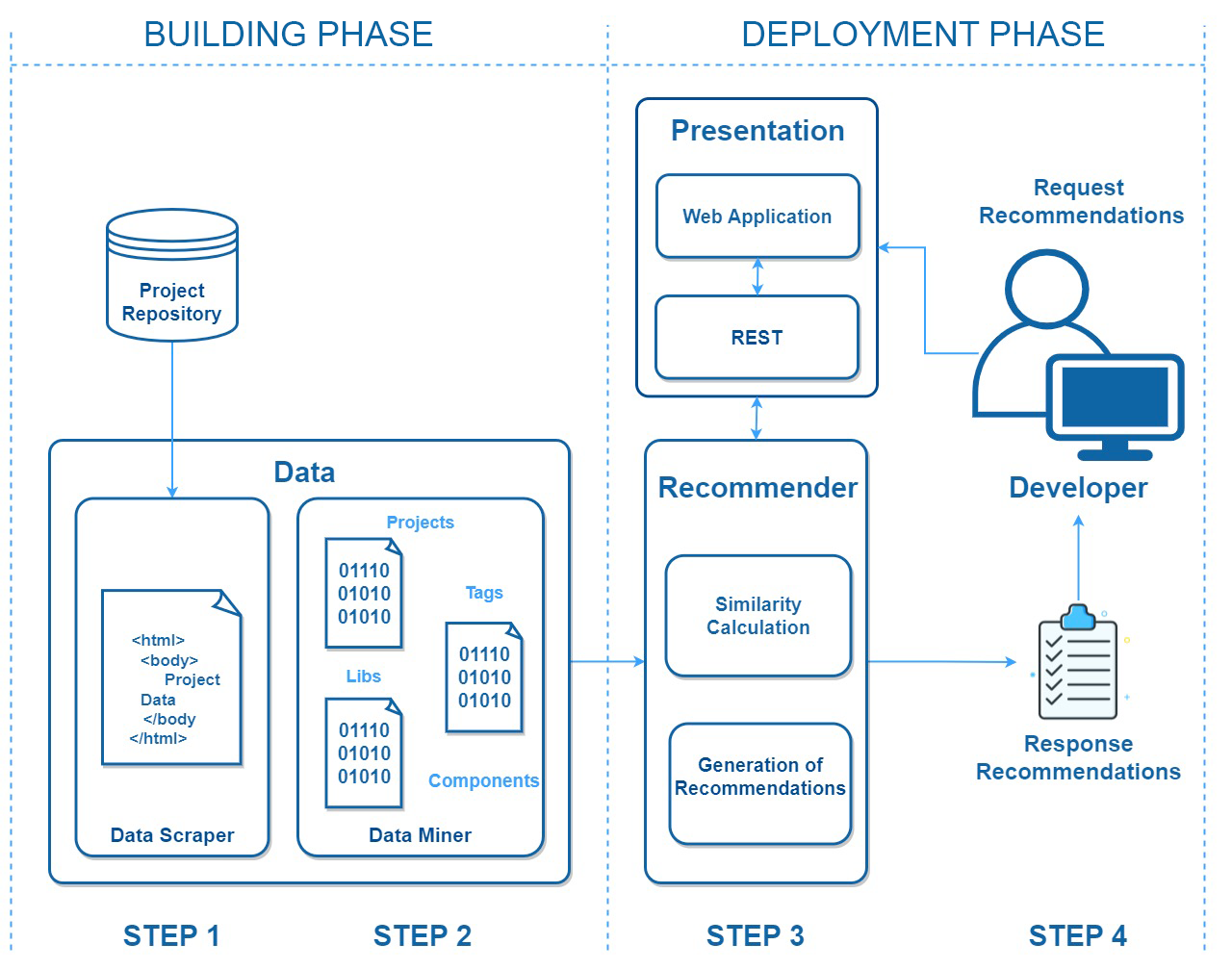}
    \caption{\RD architecture.}
    \label{fig:ResyduoOverView}
\end{figure}

%The architecture of ReSyDuo is Divided into three modules: Data Module, Recommender Module, and Presentation Module. The modules %have been identified by a separation of concern as follows:
%\begin{itemize}[noitemsep]
%\item[--]The concern of the Data Module is to extract, mine, and transform the data;
%\item[--] The concern of the Recommender Module is the similarity calculation and computation of the recommendations;
%\item[--]The concern of the Presentation module is to provide the data to the user through a web application or via REST %interface.
%\end{itemize}

\smallskip
\textit{Step 1: Data Scraper component - }
To extract the needed data, a tailored scraper has been conceived by exploiting the MDE-based approach. First, we make use of the Beautiful Soap Python library \footnote{\url{https://beautiful-soup-4.readthedocs.io/en/latest/}} to extract the available data from all the projects belonging to \PH. 
In particular, we represent the metadata available on the repository using a data model as shown in Figure~\ref{fig:ArduinoProjectModel}. The \textit{ProjectInfo} class contains information about the description, the project title, and the URL. The \textit{\cready{Statistic}} indicates the community's quality and approval of the project. Users use \textit{Tag}s to categorize projects. Finally, there are \textit{Component} and \textit{SourceCode} data. This data model aims to capture both semantic and technical information so that relationships between that information can be identified during the data mining process. %This component eventually stores the projects' metadata JSON files together with the C source code files.
In the next step, three different projections of such conceptualization will be analyzed.

\smallskip
\textit{Step 2: Data Miner component - }
Once the data have been downloaded and represented by the data model, the \texttt{Data Miner} component encodes the obtained information at three different projections \ie \textit{T}, \textit{P}, and \textit{L}.
By querying the data model depicted in Figure~\ref{fig:ArduinoProjectModel}, \textit{T}, \textit{P}, and \textit{L} represent the projections of Arduino projects and components, Arduino projects and 
tags, and components and source code, respectively.
It is worth mentioning that all the data are represented as matrices. 
The rationale behind this choice is that \textit{i)} this encoding preserves the semantic relations among the three mentioned component and \textit{ii)} the underpinning CF-based algorithm requires such an encoding scheme. For instance, to support the \typeI recommendations, this component extracts the matrix represented in Table~\ref{tab:CutOff}.
%\begin{figure}
%    \centering
%    \begin{equation}    
%    \begin{pmatrix}
%1 & 2 & 3 \\
%a & b & c
%\end{pmatrix}
%\end{equation}    
%\caption{Caption}
%\label{fig:matrix}
%\end{figure}
%
%Similarly, the \texttt{Data Miner} component builds two additional matrix, \ie \typeII and \typeIII matrixes depicted in Figure \ref{} and \ref{} respectively.
We preprocessed the raw data we scraped from ProjectHub to remove possible noise. In particular, cut-off values have been used to remove items that are very rare. For example, in project-component data,  a cut-off allows removing components that are used by a few projects. The cut-off can also be used to remove a project that contains few components. %In general, this technique can be applied to every type of data described in Sections \ref{ProjectComponentData}, \ref{TagComponentData} and \ref{ComponentLibraryData}.

\begin{table}
    \centering
    \footnotesize
    \begin{tabular}{|l|c|c|c|c|c|c|c|c|c|}
    \hline
    &\textbf{C1}  & \textbf{C2} & \textbf{C3}  &\textbf{C4}   &\textbf{C5}  &\textbf{C6}  &\textbf{C7}  &\textbf{C8}\\ \hline
    P1 & 0 & 1 & 0 & 0 &1 &1 &\cellcolor{red!20}0  &\cellcolor{red!20}0\\ \hline
    P2 & 0 & 0 & 1 &1 & 0 & 0 &\cellcolor{red!20} 1 &\cellcolor{red!20}1\\ \hline
    P3 & 0 & 0 & 0 &0 & 1 & 1 & \cellcolor{red!20}0 &\cellcolor{red!20}0\\ \hline
    P3 & 0 & 0 & 0 &0 & 1 & 1 & \cellcolor{red!20}0 &\cellcolor{red!20}0\\ \hline
    P4 & 0 & 1 & 1 &1 & 0 & 0 &\cellcolor{red!20} 0 &\cellcolor{red!20}0\\ \hline
    P5 & 1 & 1 & 0  & 0& 1 & 1 &\cellcolor{red!20} 0 &\cellcolor{red!20}0\\ \hline
    P6 & 1 & 0 & 0 &1 & 1 & 0 & \cellcolor{red!20}1 &\cellcolor{red!20}0\\ \hline
    P7 & 1 & 1 & 0 &1 & 1 & 1 &\cellcolor{red!20} 0 &\cellcolor{red!20}0\\ \hline
    \cellcolor{blue!20}P8 & \cellcolor{blue!20}0 & \cellcolor{blue!20}0 & \cellcolor{blue!20}1 & \cellcolor{blue!20}0 & \cellcolor{blue!20}1 & \cellcolor{blue!20}0 &\cellcolor{purple!40} 0 &\cellcolor{purple!40}0\\ \hline
    \cellcolor{blue!20}P9 & \cellcolor{blue!20}0 & \cellcolor{blue!20}0 & \cellcolor{blue!20}1 & \cellcolor{blue!20}0 & \cellcolor{blue!20}1 & \cellcolor{blue!20}0 & \cellcolor{purple!40}0 &\cellcolor{purple!40}0\\ \hline
    \end{tabular}
    \caption{Application of cut-off values.}\label{tab:CutOff}
\end{table}
Specifically, vertical cut-off is a technique used to eliminate a set of columns from the dataset. The vertical cut-off imposes a lower bound on the number of occurrences of an item in the entire dataset: if an item occurs less than the cut-off values, it is removed from the dataset.  Horizontal cut-off is the technique used to eliminate a set of rows from the dataset. As in the case of the vertical cut-off, a threshold is set on the minimum number of items that must appear in each row. 
Table~\ref{tab:CutOff} shows a practical example of applying  3 as vertical and horizontal cut-off values. C7 and C8 columns appear 2 and 1 times, respectively then they are removed. Even P8 and P9 rows do not exceed the horizontal cut-off value, thus the two rows are removed from the dataset

\begin{figure}
    \centering
    \includegraphics[width=1\linewidth]{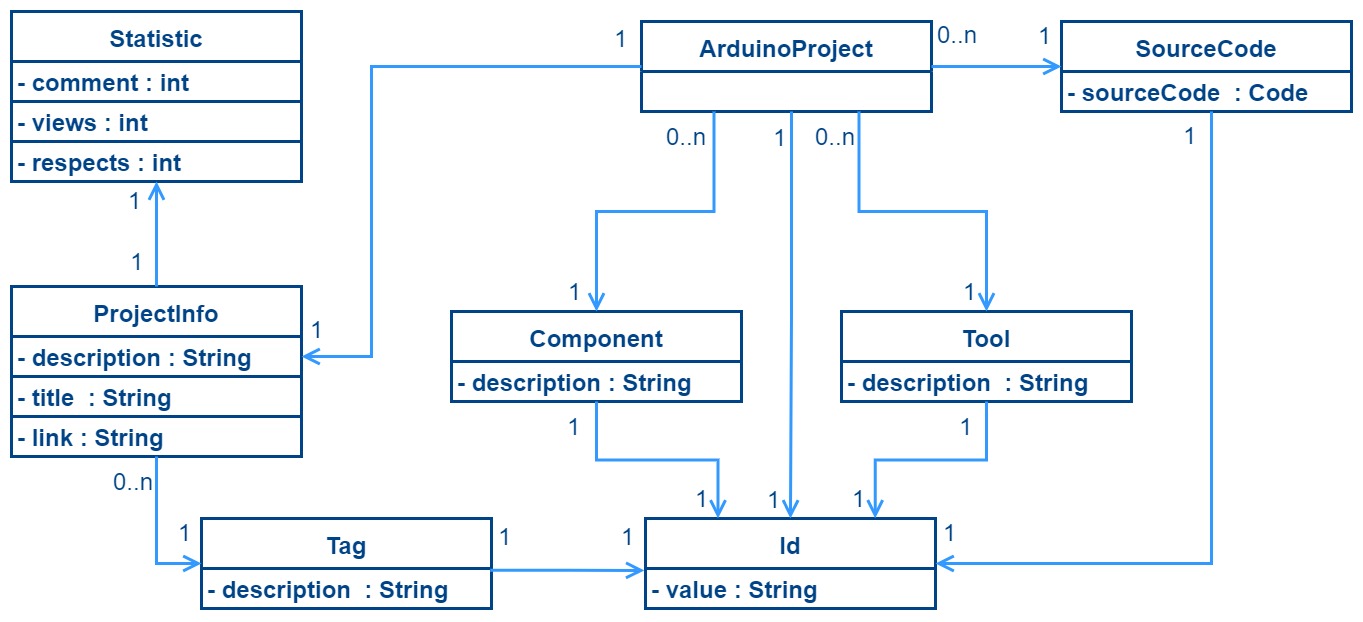}
    \caption{\PH Data Model.}
    \label{fig:ArduinoProjectModel}
\end{figure}

\begin{figure*}[h!]
    \centering
    \includegraphics[width=0.9\textwidth]{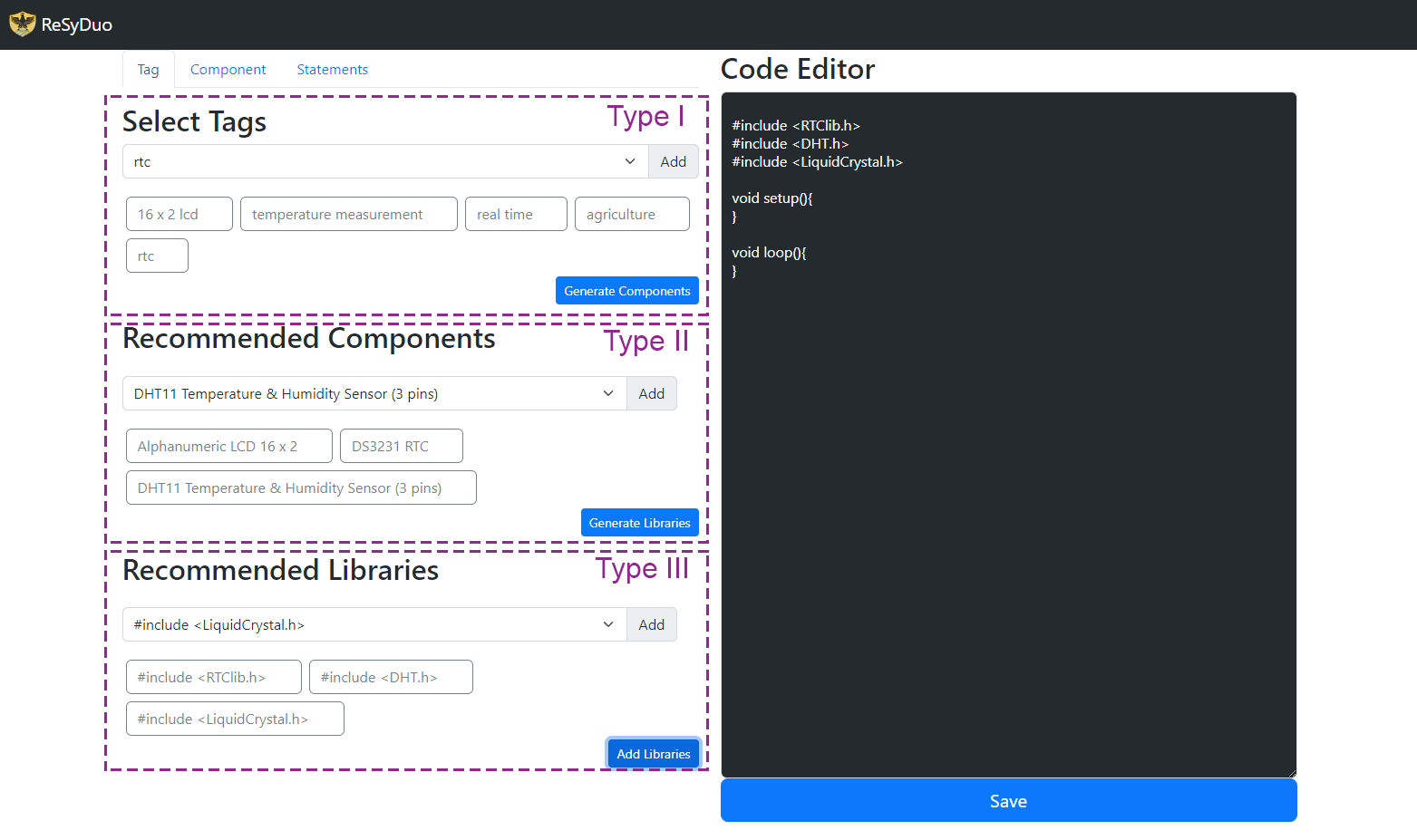}
    \caption{\RD web-based editor.}
    \label{fig:ResyduroMockup1}
\end{figure*}
\smallskip
\textit{Step 3: Recommender component - }
To recommend the abovementioned components, this component exploits a CF-based algorithm provided by the Surprise Python library \cite{hug2020surprise}. Generally speaking, a CF system is composed of \textit{users} who buys \textit{items} attributes and 
\textit{rating} it based on her experience and perceived value~\cite{chen2017performance}
%Therefore, a \textit{rating} is the association of a user and an item through a 
%value in a given unit (usually in scalar, binary, or unary form).
%The set of all ratings of a given user is also known as a \textit{user 
%profile}~\cite{chen2017performance}.
The set of all ratings given in a system by existing users can be represented in a so-called \textit{rating matrix}, where a row represents a user and a column represents an item. At this point, the system exploits a \textit{similarity function} $sim_{usr}(u_a, u_j)$ computes the \textit{weight} of the active user profile $u_a$ against each of the user profiles $u_j$ in the system. In the scope of the paper, we exploit the \PH data model to obtain three different types of matrices that capture different mutual relationships, \ie \textit{project-component}, \textit{tag-component}, and \textit{component-library}.  
Concerning the algorithm, we optimize the implementation of the well-known K-NN algorithm \cite{1053964} provided by Surprise by adopting the grid search strategy as discussed in Section \ref{sec:evaluation}. 
In addition, \RD stores the computed similarity matrixes in a dedicated static knowledge base, thus reducing the training time for existing projects.

% INIZIO SOTTOSOTTOSEZIONE

\smallskip
\textit{Step 4: Presentation component - }
Figure \ref{fig:ResyduroMockup1} depicts the presentation module that exploits the Flask framework\footnote{\url{https://flask.palletsprojects.com/en/2.3.x/}} to retrieve the abovementioned Arduino artifacts. 
\cready{At the early stage of development, the user can search for hardware components by relying on tags, \ie \typeI recommendations. In particular, the system generates an initial set of components that can be used to enable the next step, \ie \typeII recommendations. In this case, \RD makes use of the project similarity matrix to find the most similar projects given the developer's one as input. 
Once the project under development has been populated with the hardware components, the \RD interface can be used to request the corresponding software libraries needed to develop the actual Arduino project, \typeIII recommendations. Furthermore, the presentation module is equipped with a code editor view that allows the user to save and download the source code written on top of the recommended components.}

\section{Evaluation materials}
\label{sec:evaluation}

%This section presents the ReSyDuo evaluation methodology. First, we identify three research questions to assess each type of recommendation in Section \ref{sec:rqs}. Afterward, Section \ref{sec:strategy} and  presents the methodology used for the evaluation, including the metrics and the grid-search strategy. Finally, six datasets were extracted and discussed in Section \ref{sec:datasets}. 

\subsection{Research Questions} \label{sec:rqs}

To assess \RD's recommendation capabilities, we elicit the following research questions:
\begin{itemize}
\item[--] \rqfirst First, we investigate how \RD can support the developers by exploiting the tags extracted from \PH;
\item[--] \rqsecond Given the scenario when the developer set already a set of hardware components,  \RD's \typeII component can recommend additional components by relying on three different similarity functions;
\item[--] \rqthird Once the \RD has retrieved a list of hardware components using \typeI or \typeII components, the system employs the \typeIII component to map the list of hardware elements to the  
\end{itemize}

%To evaluate the effectiveness of Resyduo for RQ1, RQ2, and RQ3 has been implemented an evaluation method based on cross-validation, classification metrics, and rating evaluation metrics. The results have been analyzed in an iterative process, and several datasets have been tested to measure the effectiveness of ReSyDuo. At each iteration, a validation task has been performed and evaluated.

\subsection{Dataset}\label{sec:datasets}

%Several datasets have been produced to validate ReSyDuo, and are listed below:
%\JDR{Maybe a resuming table saves a lot of space.}
\begin{table}[h!]
    \centering
    \footnotesize    
    \begin{tabular}{|c|c|c|r|r|}
    \hline
         Dataset & Rec. Type & Cut-off & $1^{st}$ dim  &  $2^{nd}$ dim    \\ \hline
          $T$ & Type I & 1-1 & 3,137 (T) & 11,645 (C) \\ \hline
         $T^*$ & Type I & 5-5 & 1,706 (T) & 1,975 (C) \\ \hline
         
         $P$ & Type II & 1-1  & 5,547 (P) & 1,768 (C) \\ \hline
         $P^*$ & Type II & 5-5  & 1,262 (P) & 447 (C) \\ \hline

         $L$ & Type III & 1-1 & 5,346 (C) & 1,802 (L) \\ \hline
         $L^*$ & Type III & 5-5 & 1,861 (C) & 1,237 (L) \\ \hline
    \end{tabular}
    \caption{Dataset considered in the evaluation}
    \label{tab:datasets}
\end{table}

Starting from the initial raw data collected by means of the \textit{Data Miner component}, three different datasets have been excerpted, \ie $T$, $P$, and $L$. In addition, we extract an additional set of three datasets by applying the cutoff configurations as discussed in Section \ref{sec:approach}. It is worth mentioning that each type of dataset has been conceived for each type of recommendation, \ie \typeI, \typeII, and \typeIII.

Table \ref{tab:datasets} summarizes the considered dataset by specifying the recommendation type, the applied cut-off, and the two dimensions in terms of recommendation items that have been considered to obtain the rating matrices, \ie tags \textit{(T)}, projects \textit{(P)}, hardware components \textit{C}, and libraries \textit{(L)}. As discussed in Section \ref{sec:approach}, the cut-off values refer to the rows and columns of the rating matrix.  In the scope of the paper, we experiment with two different configurations, \ie (1,1) and (5,5). Applying the former preserves the structure of the matrix since all the items appear at least one time. In the latter, we rid of unfrequent items, \ie only the ones that appear five times in a row or column are preserved.     

The features of the selected datasets are the following:

\begin{itemize}
    \item $T$ and $T^*$ datasets have been extracted to support \typeI recommendations, \ie tag-based recommendations of hardware components. $T$ is obtained by applying the (1,1) configuration and it is composed of 3,137 different tags and 11,645 hardware components. Starting from this, we obtain $T^*$ by removing unfrequent times by applying the (5,5) cutoff. Such a process end with 1,706 tags and 1,975 components;
    \item To enable \typeII recommendations, \RD has to be trained by using \AD projects and their hardware components. Therefore, the initial dataset $P$ contains 5,547 projects and 1,768 components. To reduce the sparsity of the dataset, we apply the second cut-off that reduces the total number of projects and components to 1,262 and 447 respectively;
    \item Finally, \RD can suggest software libraries by relying on $L$ and $L^*$ datasets. Following the same process, we first consider 5,346 hardware components and 1,802 by applying the first cut-off values. Meanwhile, $L^*$ is composed of 1,861 components and 1,237 libraries by adopting the (5,5) cut-off configuration.      
\end{itemize}

\subsection{Methodology and Metrics} \label{sec:metrics}
To evaluate the accuracy of \RD, we perform the ten-fold cross-validation \cite{RefaeilzadehTL09} by splitting each of the proposed datasets into training and testing parts as shown in Figure~\ref{fig:CrossValidation}. %Such a process has been conducted for each fold to avoid any bias in performing the grid search optimization.  
\cready{In particular, the extracted datasets, \ie \textit{T}, \textit{T$_*$}, \textit{P}, \textit{P$_*$}, \textit{L}, and \textit{L$_*$}, are split into train and test parts composed of 90\% and 10\% respectively. Such a process has been conducted for each fold to avoid any bias in performing the grid search optimization.}

\begin{figure}[h!]
    \centering
    \includegraphics[width=0.8\linewidth]{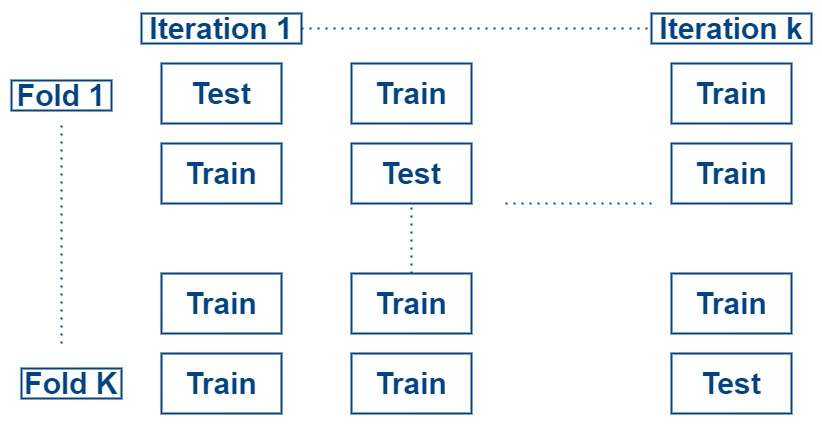}
    \caption{Ten Fold Cross Validation.}
    \label{fig:CrossValidation}
\end{figure}

To evaluate both the accuracy and the diversity of the proposed recommendations, we considered two different types of metrics, \ie \textit{accuracy} and \textit{ranking} ones.

\noindent
\textbf{Accuracy metrics -}
\cready{Given the set of recommendations $R$, we consider \textit{true positives (Tp)} as the set of the items that are correctly recommended while the items which actually do not belong to the ground-truth data are classified as \textit{false positives (Fp)}, Finally, \textit{false negatives (Fn)} are the elements that should be present in the recommended ones, but they are not. By relying on those definitions, we compute the accuracy metrics as follows:}

\noindent
\textit{Success rate:} Given a set of $R$ testing repositories, this metric measures the rate at which a recommendation engine can return at least 1 correct tag for each repository \emph{r}.
\begin{equation} \label{eqn:SuccessRate}
Success\ rate=\frac{ count_{r \in R}(   Tp  > 0 ) }{\left | R \right |} \times 100\%
\end{equation}
\noindent
where the function \textit{count()} is a boolean function that returns 1 for each true positive.

\textit{Precision:} the metric measures the rate of correct items over the entire set of recommended items:
\begin{equation}
    Precision = \frac{Tp}{Tp + Fp}
\label{eq:precision}
\end{equation}

\textit{Recall:} the ratio of the user's topics appearing in the N recommended items:
\begin{equation}
    Recall = \frac{Tp}{Tp + Fn}
\label{eq:recall}
\end{equation}

%If the recommendations produced are the ratings that users would assign to objects that have not yet been rated, then two metrics useful for evaluating this aspect are root-mean-squared-error (RMSE) and mean-absolute-error (MAE) \cite{RSSE_BOOK_2014, sarwar2001item, Schafer_2007}.
%Analyzing the predictions of a test set whose rating values are already known, the difference between the known value and the predicted value would be the error.

%The difference is given by; this metric allows us to understand if the system has overestimated or underestimated the predictions; in the first case, the error will be positive in the second negative as mentioned by Bellogin in \cite{Bellogin_2011}.\\

\noindent
\textbf{Ranking metrics -}
Assuming that an object $u$ has been rated at least by one user, the Root Mean Squared Error (RMSE) is calculated as follows:
\begin{equation}
\label{eqn:rmse}
RMSE(T) = \sqrt{\frac{\sum_{(u,i)\in T} (r_{ui}' - r_{ui})^2}{N}}
\end{equation}

where  $(r_{ui})$ and $(r_{ui}')$ are the actual rating and predicted one respectively.  
\newline
Meanwhile, the Mean Absolute Error calculates the absolute value of the average of the differences as follows:
\newline
\begin{equation}
\label{eqn:mae}
MAE(T) = \frac{\sum_{(u,i)\in T} |r_{ui}' - r_{ui}|}{N}
\end{equation}
\newline

where $T$ is the test set of possible user-item pairs and $N$ is the cardinality of the set of ratings. In this paper, we employ the normalized version of  RMSE and MAE according to the rating scale. 

%In MAE, the differences are weighted in the same way, while in the RMSE bigger differences are penalized more than the smaller ones because the squares are calculated before calculating the mean.\\

%For both metrics, a low value indicates a good degree of correctness of the RS, and usually, the RMSE is greater than or equal to the MAE.\\

%\subsection{Similarity measures}\label{sec:similarity}

%\begin{equation}
%\label{eqn:10}
%MSD(u,n) = \frac{1}
%{|CR_{u,n}|} \cdot \sum_{i \in CR_{u,n}}(r_{ui}-\overline{r}_u)^2\\
%\end{equation}
%So the \textit{MSD} Similarity is defined as:
%\begin{equation}
%\label{eqn:11}
%simMSD(u,n) = \frac{1}
%{MSD(u,v)+1}\\
%\end{equation}

%All the histograms show long tails distribution trend. In which a small set of items appear in many projects, this set dominates the distribution. The set of the dominant items creates a connection between the different projects and allows us to calculate the similarity between them. The objects in the long tail are still useful because they diversify the projects, and it is reasonable to keep them in the dataset.

\subsection{Hyperparameter selection} \label{sec:grid}
Predictive systems require hyperparameter optimization to automatically set parameter combinations for optimal model or algorithm performance, reducing human effort and enhancing experiment reproducibility \cite{automatedML_2019}. Among the others, the grid search strategy \cite{doi:10.1080/1206212X.2021.1974663} involves specifying finite parameter values and evaluating the Cartesian product, but it suffers from the dimensionality problem due to exponential evaluations as parameter count increases. To avoid any bias in the implementation, we exploit the Surpise dedicated function to optimize the selected algorithm \footnote{\url{https://tinyurl.com/2u6frv27}}. 

In the scope of the evaluation, we experiment with three different similarity functions, \ie \textit{Pearson Correlation. Cosine, and Mean Squared Difference (MSD)}. %Given $u$ and $i$ two users, the \textit{co-rating} is the set of items rated by both users $u$ and $i$ and is denoted by $CR_{u,n}$.
The \textit{Pearson Correlation} \cite{Schafer_2007} coefficient is a measure of dependence between two paired random vectors. %computed as follows:
The \textit{Cosine} similarity function \cite{7577578} is based on the calculation of the spatial distance between the ratings of all user pairs. %:
The \textit{Mean Squared Difference}~\cite{bickel2015mathematical} assesses the average squared difference between the observed and predicted values.

%Finally, different neighbors values have been considered, \ie 5, 10, 20. 
According to the selected technique, we consider if the similarity has been computed between users or items as stated in \textit{CF technique} column. Finally, the \textit{Min Common Item} and \textit{K-Neighbor} columns indicate the minimum rated common item between users and the neighbor set's cardinality respectively. 
 
\begin{table}[h!]
    \centering
    \scriptsize
    \begin{tabular}{|c|c|c|c|c|}
    \hline
    \textbf{Similarity Function}  & \textbf{CF technique}& \textbf{Min Common Items}&\textbf{ K-Neighbor} \\ \hline
     \cellcolor{green!20}MSD & Item based & \cellcolor{green!20}1 & 5 \\ \hline
    Cosine & \cellcolor{green!20}User based & 5 &10 \\ \hline
    Pearson & - & 20 & \cellcolor{green!20}20 \\ \hline
    \end{tabular}
    \caption{Results of the grid search optimization.}
    \label{tab:GridSearchTable}
\end{table}

Table \ref{tab:GridSearchTable} summarizes the results obtained by running the grid search algorithm. In particular, the cells highlighted in green indicate that the best configuration includes the following hyperparameters: Sim\ =\ MSD,\ User\_Based\ =\ True,\ Min\_Support = 1,\ K\ =\ 20. Such a configuration has been employed to assess the \RD's overall performances in the next section.

%The Similarity Function column matches the type of function used to calculate the similarity matrix (see formula \ref{eqn:8}, \ref{eqn:9}, \ref{eqn:11}). The User-Based column indicates if the similarity has been computed between Users or Between Items, Min Common Item indicates the minimum rated common item between users, and K-Neighbor indicates the Neighbor set's cardinality.
%The grid search result is indicated by the green cells of the table. The optimization process has been tested with a tenfold cross-validation with the K-NN with means algorithm.
%So, the following hyperparameters: Sim\ =\ MSD,\ User\_Based\ =\ True,\ Min\_Support = 1,\ K\ =\ 20.

\section{Results}
\label{sec:results}
In this section, we aim at assessing the proposed approach by considering the three types of provided recommendations. To this end, we employ the evaluation process described in Section \ref{sec:evaluation} by employing the mined datasets, the identified set of metrics, and the elicited hyperparameters using the grid search strategy. 
In particular, Sections \ref{sec:RQ1}, \ref{sec:RQ2}, and \ref{sec:RQ3} present the results obtained for the \typeI, \typeII, and \typeIII recommendations. 
We eventually discuss the limitations and the threats to validity in Section \ref{sec:threats}.

%The metrics that had the most significant weight in the evaluation of the performance are precision, recall, and success rate, while the MAE and RMSE were used to measure the precisions of predicted ratings e find a correlation between the classification metrics and the rating evaluation metrics.\\

% INIZIO SOTTOSEZIONE
\subsection{\rqfirst}\label{sec:RQ1}

%By adopting two different cut-off settings, \ie <1,1>  and <5.5> as vertical and horizontal cut-off values, we evaluated the performance of \RD. 
To answer this question, we consider the first pair of datasets extracted from the \PH repository, \ie P and P$^*$ as discussed in Section \ref{sec:evaluation}.
 Table \ref{tab:p_datasets_comparison} summarizes the obtained results on the examined datasets. From the accuracy point of view, it is evident that \RD performs better when dataset P$^*$ is considered, \ie removing unfrequent items increases the overall performances. In particular, the precision and recall have been increased by 10\% thus reaching 0.203 and 0.227 on average. Similarly, the success rate grows up to 0.271 by considering the second cut-off configuration, \ie (5,5).

 %and the success rate has risen to 27\%. 

 The positive effect of the cut-off configuration holds by considering the raking metrics, \ie,  the MAE and RMSE value passes from 0.006 to 0.033 and from 0.0048 to 0.118 respectively. Nonetheless, the observed scores are dramatically low, meaning that on average the elements suggested by \RD are not proper for the current project.  Such low results can be explained by considering that we didn't curate the mined dataset apart from removing unfrequent items.   

% \CDS{è vero? Controllare!!}
% Such low results can be explained by considering the fact that we provide \textit{diverse} recommendations, \ie we avoid recommending always popular items. 

 %and the average RMSE at 0.12 in a rating range from 0 to 1. The evaluation of the rating between dataset P and P$^*$ is discordant with the evaluation of classification metrics such as precision, recall, and success rate. In P the classification metrics are worse than those of rating evaluation are better. On the opposite, in P$^*$ the classification metrics return better results than the rating metrics.

\begin{tcolorbox} \textbf{Answer to RQ$_1$:} Considering \typeI recommendations, the cut-off configuration contributes to increasing the overall performances, even though the positive effect is marginal when raking metrics are considered.  
\end{tcolorbox}

\subsection{\rqsecond}\label{sec:RQ2}

%To answer this research question, two datasets T1 and T2 were produced, and  tested. The characteristics of these datasets are shown in detail in \ref{Dataset}.

\begin{table*}[]
\centering
\begin{tabular}{|l|llllllllll|}
\hline
                      & \multicolumn{2}{c|}{\textbf{Precision}}                               & \multicolumn{2}{c|}{\textbf{Recall}}                                  & \multicolumn{2}{c|}{\textbf{Success rate}}                            & \multicolumn{2}{c|}{\textbf{MAE}}                                     & \multicolumn{2}{c|}{\textbf{RMSE}}                                    \\ \hline
\textbf{(v,h) cutt-off} & \multicolumn{1}{l|}{\textbf{(1,1)}} & \multicolumn{1}{l|}{\textbf{(5,5)}} & \multicolumn{1}{l|}{\textbf{(1,1)}} & \multicolumn{1}{l|}{\textbf{(5,5)}} & \multicolumn{1}{l|}{\textbf{(1,1)}} & \multicolumn{1}{l|}{\textbf{(5,5)}} & \multicolumn{1}{l|}{\textbf{(1,1)}} & \multicolumn{1}{l|}{\textbf{(5,5)}} & \multicolumn{1}{l|}{\textbf{(1,1)}} & \multicolumn{1}{l|}{\textbf{(5,5)}} \\ \hline
\textbf{Fold 1}       & 0.108                            & 0.221                            & 0.101                             & 0.236                            & 0.116                            & 0.284                            & 0.006                            & 0.034                            & 0.047                            & 0.120                              \\ \cline{1-1}
\textbf{Fold 2}       & 0.104                            & 0.206                            & 0.099                            & 0.235                            & 0.114                            & 0.272                            & 0.006                            & 0.034                            & 0.048                            & 0.121                            \\ \cline{1-1}
\textbf{Fold 3}       & 0.111                            & 0.195                            & 0.104                            & 0.212                             & 0.121                             & 0.253                            & 0.006                            & 0.033                            & 0.048                            & 0.117                           \\ \cline{1-1}
\textbf{Fold 4}       & 0.111                           & 0.201                            & 0.102                            & 0.218                            & 0.120                            & 0.268                            & 0.006                            & 0.033                            & 0.048                            & 0.116                            \\ \cline{1-1}
\textbf{Fold 5}       & 0.109                            & 0.196                            & 0.101                            & 0.215                            & 0.119                            & 0.257                            & 0.006                            & 0.034                            & 0.048                            & 0.119                            \\ \cline{1-1}
\textbf{Fold 6}       & 0.108                            & 0.210                            & 0.100                            & 0.239                            & 0.116                            & 0.284                            & 0.006                            & 0.034                             & 0.048                            & 0.118                            \\ \cline{1-1}
\textbf{Fold 7}       & 0.107                            & 0.193                            & 0.099                            & 0.223                            & 0.115                            & 0.263                            & 0.006                            & 0.034                            & 0.048                        & 0.118                            \\ \cline{1-1}
\textbf{Fold 8}       & 0.106                            & 0.197                            & 0.098                            & 0.225                            & 0.114                            & 0.266                            & 0.006                            & 0.034                            & 0.047                            & 0.119                            \\ \cline{1-1}
\textbf{Fold 9}       & 0.109                            & 0.222                            & 0.100                            & 0.244                             & 0.118                            & 0.296                            & 0.006                            & 0.034                            & 0.047                            & 0.120                              \\ \cline{1-1}
\textbf{Fold 10}      & 0.107                            & 0.192                             & 0.100                               & 0.227                            & 0.116                            & 0.264                            & 0.006                            & 0.034                            & 0.047                            & 0.117                          \\ \hline
\textbf{Avg.}         & 0.108                             & \textbf{0.203}                           & 0.100                           & \textbf{0.227}                           & 0.117                           & \textbf{0.271}                           & 0.006                            & \textbf{0.033}                           & 0.048                           &  \textbf{0.118}                           \\ \hline
\end{tabular}    
\caption{\RD's results considering P and P$^*$} 
%- Hyperparameter: Sim\ =\ MSD,\ User\_Based\ =\ True,\ Min\_Support = 1,\ K\ =\ 20.}
\label{tab:p_datasets_comparison}
\end{table*}

\begin{table*}[]
\centering
\begin{tabular}{|l|llllllllll|}
\hline
                      & \multicolumn{2}{c|}{\textbf{Precision}}                               & \multicolumn{2}{c|}{\textbf{Recall}}                                  & \multicolumn{2}{c|}{\textbf{Success rate}}                            & \multicolumn{2}{c|}{\textbf{MAE}}                                     & \multicolumn{2}{c|}{\textbf{RMSE}}                                    \\ \hline
\textbf{(v,h) cutt-off} & \multicolumn{1}{l|}{\textbf{(1,1)}} & \multicolumn{1}{l|}{\textbf{(5,5)}} & \multicolumn{1}{l|}{\textbf{(1,1)}} & \multicolumn{1}{l|}{\textbf{(5,5)}} & \multicolumn{1}{l|}{\textbf{(1,1)}} & \multicolumn{1}{l|}{\textbf{(5,5)}} & \multicolumn{1}{l|}{\textbf{(1,1)}} & \multicolumn{1}{l|}{\textbf{(5,5)}} & \multicolumn{1}{l|}{\textbf{(1,1)}} & \multicolumn{1}{l|}{\textbf{(5,5)}} \\ \hline
\textbf{Fold 1}       & 0.203                            & 0.293                            & 0.133                             & 0.259                            & 0.224                            & 0.402                            & 0.006                            & 0.037                            &  0.216                           & 0.451                              \\ \cline{1-1}
\textbf{Fold 2}       & 0.188                            & 0.278                            & 0.126                            & 0.240                            & 0.208                            & 0.372                            & 0.006                            & 0.039                            & 0.177                            & 0.569                            \\ \cline{1-1}
\textbf{Fold 3}       & 0.191                            & 0.303                            & 0.125                            & 0.253                             & 0.211                             & 0.400                            & 0.006                            & 0.038                            & 0.149                            & 0.420                           \\ \cline{1-1}
\textbf{Fold 4}       & 0.189                           &  0.292                           & 0.122                            & 0.248                            & 0.206                            & 0.383                            & 0.006                            & 0.038                            & 0.196                            & 0.521                            \\ \cline{1-1}
\textbf{Fold 5}       & 0.194                            & 0.308                            & 0.125                            & 0.257                            & 0.215                            & 0.406                            & 0.006                            & 0.037                            & 0.163                            & 0.335                            \\ \cline{1-1}
\textbf{Fold 6}       & 0.183                            & 0.294                            & 0.122                            & 0.242                            & 0.203                            & 0.388                            & 0.006                            & 0.039                             & 0.326                            & 0.837                            \\ \cline{1-1}
\textbf{Fold 7}       & 0.196                            & 0.288                            & 0.130                            & 0.248                            & 0.214                            & 0.390                            & 0.006                            & 0.040                            & 0.171                        & 0.885                            \\ \cline{1-1}
\textbf{Fold 8}       & 0.186                            & 0.293                            & 0.120                            & 0.250                            & 0.206                            & 0.398                            & 0.006                            & 0.038                            & 0.158                            & 0.551                            \\ \cline{1-1}
\textbf{Fold 9}       & 0.189                            & 0.299                            & 0.124                            & 0.258                             & 0.208                            & 0.398                            & 0.006                            & 0.038                            & 0.158                            & 0.423                              \\ \cline{1-1}
\textbf{Fold 10}      & 0.192                            & 0.297                             & 0.128                               & 0.255                            & 0.214                            & 0.399                            & 0.006                            & 0.039                            & 0.130                            & 0.622                          \\ \hline
\textbf{Avg.}         & 0.190                             & \textbf{0.290}                           & 0.130                           & \textbf{0.250}                           & 0.210                           & \textbf{0.390}                           & 0.006                            & \textbf{0.040}                           & 0.180                           & \textbf{0.560}                           \\ \hline
\end{tabular}    
\caption{\RD's results considering T and T$^*$} 

%- Hyperparameter: Sim\ =\ MSD,\ User\_Based\ =\ True,\ Min\_Support = 1,\ K\ =\ 20.}
\label{tab:t_datasets_comparison}
\end{table*}

%\begin{table*}[h!]
%\footnotesize
%\centering
%\begin{tabular}{|l|c|c|c|c|c|}
%\hline
%\multicolumn{6}{|c|}{$Dataset\ T1:\ Sim\ =\ MSD,\ User\_Based\ =\ True,\ Min\_Support = 1,\ K\ =\ 20$} \\
%\cline{1-6}
%\multicolumn{2}{|c|}{Cut-Off (row-col)} & \multicolumn{2}{c|}{File Size} & \multicolumn{2}{c|}{Dimension (row,col)}\\
%\cline{1-6}
%\multicolumn{2}{|c|}{1,1} & \multicolumn{2}{c|}{1462.1169} & \multicolumn{1}{c|}{3137} & \multicolumn{1}{c|}{11645}\\
%\cline{1-6}
%\multicolumn{1}{|c|}{fold} & Precision & Recall  & Success rate &MAE &RMSE\\
%
%\hline
%Fold1 &0.2036 &0.1336  &0.2241 &0.0061&0.2163\\
%Fold2 &0.1887 &0.1261  &0.2085 &0.0061&0.1774\\
%Fold3 &0.1914 &0.1252  &0.2117 &0.006&0.1496\\
%Fold4 &0.1895 &0.1228  &0.2069 &0.0061&0.1967\\
%Fold5 &0.1949 &0.1259  &0.2155 &0.0061&0.1637\\
%Fold6 &0.1835 &0.122  &0.2031 &0.0064&0.3263\\
%Fold7 &0.1965 &0.1302  &0.2149 &0.0061&0.1711\\
%Fold8 &0.1865 &0.1209  &0.2062 &0.0061&0.1584\\
%Fold9 &0.1899 &0.1244  &0.2085 &0.006&0.1588\\
%Fold10 &0.1923 &0.1283  &0.2149 &0.006&0.1302\\
%\cline{1-6}Average &0.19 &0.13 &0.21 &0.01&0.18\\
%\hline
%\end{tabular}
%\caption{Analysis of the Predicted Results for the Dataset T1.}\label{tab:resultP1}
%\end{table*}

Similar to the previous research question, Table \ref{tab:t_datasets_comparison} compares the results obtained by \RD's in terms of Type-II recommendations by relying on two different datasets, \ie T and T*. As for the \typeI recommendations, we adopt such data curation to reduce the sparsity of the dataset.
 
 %Dataset T1, similarly to P1 has been used to measure performance with a very sparse dataset; it can be considered as a stress test to validate the recommendations of Type II. From the results in table \ref{tab:resultP1}, we can deduce that all the portions of the dataset have a similar distribution; 
 
 Overall, the results are better compared to the \typeI, \ie \RD archives better performance for all the considered metrics. In particular, the success rate achieves the best results among the accuracy metrics by reaching 0.390 as an average value. The same positive effect can be observed on both the considered raking metrics, \ie the MAE and RMSE are increased by 0.011, and 0.442. Even though the improvements are marginal, the conducted data curation succeded in improving the overall performances. Concerning the cut-off configuration, the T* dataset confirms the findings of the first research question, \ie the applied quality filter improves also the \typeII recommendations.

\begin{tcolorbox} \textbf{Answer to RQ$_2$:} By applying additional data curation, \RD \typeII recommendations overcome the \typeI ones in terms of all considered metrics albeit the improvements are marginal. 
\end{tcolorbox} 

\subsection{\rqthird}\label{sec:RQ3}

To support \typeIII recommendations, two datasets L and L$^*$ (containing relationships between hardware components and the corresponding software components) were produced following the same process adopted for the abovementioned recommendations.

%\begin{table*}[h!]
%\footnotesize
%\centering
%\begin{tabular}{|l|c|c|c|c|c|c|c|c|c|}
%\hline
%\multicolumn{10}{|c|}{Comparison of Predicted Results for Datasets %T1 and T2} \\
%\hline
%\multicolumn{5}{|c|}{\textbf{T1 Dataset}} & \multicolumn{5}{c|}%{\textbf{T2 Dataset}} \\
%\hline
%
%\multicolumn{1}{|c|}{fold} & Precision & Recall  & Success rate & %MAE & \multicolumn{1}{c|}{fold} & Precision & Recall  & Success rate %& MAE \\
%\hline
%0.2036 &0.1336  &0.2241 &0.0061&0.2163 & Fold1 &0.2939 &0.2599  %&0.4027 &0.0377&0.4518 \\
%0.1887 &0.1261  &0.2085 &0.0061&0.1774 & Fold2 &0.2783 &0.2402  %&0.3728 &0.0396&0.5693 \\
%0.1914 &0.1252  &0.2117 &0.006&0.1496 & Fold3 &0.3039 &0.2535  %&0.4004 &0.0384&0.4203 \\
%0.1895 &0.1228  &0.2069 &0.0061&0.1967 & Fold4 &0.2921 &0.2488  %&0.3839 &0.0385&0.5215 \\
%0.1949 &0.1259  &0.2155 &0.0061&0.1637 & Fold5 &0.3083 &0.2579  %&0.4068 &0.0377&0.3356 \\
%0.1835 &0.122  &0.2031 &0.0064&0.3263 & Fold6 &0.2942 &0.2421  %&0.3886 &0.0399&0.8371 \\
%0.1965 &0.1302  &0.2149 &0.0061&0.1711 & Fold7 &0.2888 &0.2489  %&0.3904 &0.0409&0.8854 \\
%0.1865 &0.1209  &0.2062 &0.0061&0.1584 & Fold8 &0.2937 &0.2503  %&0.3986 &0.0388&0.551 \\
%0.1899 &0.1244  &0.2085 &0.006&0.1588 & Fold9 &0.2993 &0.2585  %&0.3986 &0.0384&0.4237 \\
%0.1923 &0.1283  &0.2149 &0.006&0.1302 & Fold10 &0.2971 &0.2557  %&0.3998 &0.0395&0.622 \\
%\hline
%Average &0.19 &0.13 &0.21 &0.01&0.18 & Average &0.29 &0.25 &0.39 %&0.04&0.56 \\
%\hline
%\end{tabular}
%\caption{Comparison of Predicted Results for Datasets T1 and T2.}
%\end{table*}

\begin{table*}[]
\centering
\begin{tabular}{|l|llllllllll|}
\hline
                      & \multicolumn{2}{c|}{\textbf{Precision}}                               & \multicolumn{2}{c|}{\textbf{Recall}}                                  & \multicolumn{2}{c|}{\textbf{Success rate}}                            & \multicolumn{2}{c|}{\textbf{MAE}}                                     & \multicolumn{2}{c|}{\textbf{RMSE}}                                    \\ \hline
\textbf{(v,h) cutt-off} & \multicolumn{1}{l|}{\textbf{(1,1)}} & \multicolumn{1}{l|}{\textbf{(5,5)}} & \multicolumn{1}{l|}{\textbf{1,1}} & \multicolumn{1}{l|}{\textbf{(5,5)}} & \multicolumn{1}{l|}{\textbf{(1,1)}} & \multicolumn{1}{l|}{\textbf{(5,5)}} & \multicolumn{1}{l|}{\textbf{(1,1)}} & \multicolumn{1}{l|}{\textbf{(5,5)}} & \multicolumn{1}{l|}{\textbf{(1,1)}} & \multicolumn{1}{l|}{\textbf{(5,5)}} \\ \hline
\textbf{Fold 1}       & 0.209                            & 0.237                            & 0.203                             & 0.527                            & 0.234                            & 0.586                            & 0.011                            & 0.057                            &  0.230                           & 0.605                              \\ \cline{1-1}
\textbf{Fold 2}       & 0.210                            & 0.250                            & 0.203                            & 0.536                            & 0.236                            & 0.599                            & 0.012                            & 0.059                            & 0.339                            & 0.595                            \\ \cline{1-1}
\textbf{Fold 3}       & 0.211                            & 0.236                            & 0.204                            & 0.521                             & 0.235                             & 0.582                            & 0.011                            & 0.055                            & 0.189                            & 0.791                           \\ \cline{1-1}
\textbf{Fold 4}       & 0.209                           &  0.240                           & 0.206                            & 0.520                            & 0.238                            & 0.583                            & 0.011                            & 0.057                            & 0.168                            & 0.556                            \\ \cline{1-1}
\textbf{Fold 5}       & 0.207                            & 0.250                            & 0.202                            & 0.554                            & 0.233                            & 0.614                            & 0.011                            & 0.058                            & 0.341                            & 0.636                            \\ \cline{1-1}
\textbf{Fold 6}       & 0.208                            & 0.251                            & 0.203                            & 0.535                            & 0.233                            & 0.597                            & 0.012                            & 0.055                             & 0.364                            & 0.619                            \\ \cline{1-1}
\textbf{Fold 7}       & 0.203                            & 0.245                            & 0.193                            & 0.538                            & 0.224                            & 0.599                            & 0.011                            & 0.054                            & 0.175                        & 0.774                            \\ \cline{1-1}
\textbf{Fold 8}       & 0.210                            & 0.247                            & 0.204                           & 0.529                            & 0.236                            & 0.587                            & 0.011                            & 0.056                            & 0.221                            & 0.659                            \\ \cline{1-1}
\textbf{Fold 9}       & 0.202                            & 0.246                            & 0.194                            & 0.518                             & 0.227                            & 0.577                            & 0.011                            & 0.058                            & 0.326                            & 0.592                              \\ \cline{1-1}
\textbf{Fold 10}      & 0.208                            & 0.247                             & 0.205                               & 0.546                            & 0.234                            & 0.607                            & 0.011                            & 0.060                            & 0.194                            & 0.675                          \\ \hline
\textbf{Avg.}         & 0.210                             & \textbf{0.250}                           & 0.200                           & \textbf{0.530}                           & 0.230                           & \textbf{0.590}                           & 0.010                            & \textbf{0.060}                           & 0.250                           & \textbf{0.650}                           \\ \hline
\end{tabular}    
\caption{\RD's results considering L and L$^*$} 
%- Hyperparameter: Sim\ =\ MSD,\ User\_Based\ =\ True,\ Min\_Support = 1,\ K\ =\ 20.}
\label{tab:l_datasets_comparison}
\end{table*}

As for the previous \textbf{RQs}, the dataset L$^*$ has been extracted from the L dataset 
 and represents a curated version of the original one, \ie the degree of data homogeneity is higher. In particular, precision, recall, and success rate reaches 0.250, 0.530, and 0.590 respectively. The raking metrics are improved as well, although the results are still far from being optimal, \ie the RSME achieves O.650 with the best cut-off configuration. This can be explained by considering the nature of the dataset, \ie no quality filter has been applied during the data-gathering phase. Nonetheless, the improvements of \typeIII recommendations are encouraging.

 %The rating evaluation metrics returned low results; the average MAE stands at 0.06, and the average RMSE at 0.65 in a rating value range from 0 to 10. 
%The evaluation of the rating between datasets L1 and L2 is discordant with the evaluation of classification metrics such as precision, recall, and success rate. In L1's evaluation, the classification metrics are worse than L2, and the rating evaluation in L1 is better than L2.

 %We can confirm that ReSyDuo is more effective at recommending libraries than hardware components.

%\vspace{1cm}
%box
\begin{tcolorbox} \textbf{Answer to RQ$_3$:} \RD succeeded in providing relevant software TPLs, \ie \typeIII recommendations obtains better performance compared to \typeI and \typeII. Albeit metrics values are not exciting, the approach supports the recommendation of \iot libraries in an effective manner. 

\end{tcolorbox}

\section{Threats to validity}
\label{sec:threats}
This section discusses threats that may hamper the obtained results. Concerning \textit{internal validity}, we employ a tailored data model to collect \AD projects from \PH without applying any quality filter on them, \eg considering the number of respects or views. We mitigate this issue by adopting two different cut-off configurations on the matrices to remove unfrequent projects, tags, and libraries. Moreover, the selected KNN algorithm may lead to additional threats. To mitigate this, we adopt the grid search optimization strategy to automatically select the best hyperparameters configuration.

\textit{External validity} concerns the lack of a proper baseline. To our knowledge, no suitable approach provides the three types of recommendations supported by \RD. We conduct a well-structured quantitative evaluation by applying a grid search strategy to optimize the hyperparameters of the CF-based engine. Furthermore, we considered two different kinds of metrics, \ie accuracy, and ranking, to analyze the obtained results from different perspectives.

\section{Related works}
\label{sec:related}

\subsection{Tag-based recommendation systems}

MUDABlue \cite{kawaguchi_mudablue_2004} clusters similar software projects by exploiting the LSA technique. First, the system infers clusters by relying on a matrix-based representation of projects. Afterward, the LSA algorithm is employed to retrieve similar projects to the input one using the computed clusters.

Zuo et al. in \cite{zuo2016tag} presented a tag-aware RS based on deep neural networks. First, the users' profiles are modeled as vectors over tags. Afterward, the underpinning network retrieves latent features from the
users' tag space. The proposed system eventually aggregates the extracted features and item information to generate recommendations. The evaluation shows that the proposed approach reduces the sparsity on the two tested datasets.

%In this approach, relationships are built between user-defined tags and users. In this paper, particular attention is focused on the similarity computation problem caused by the sparsity of the relationships between tags and users. This RS is based on a neural network approach instead of a classic collaborative filtering approach. %Similarities have been found between these recommender systems and Type II recommendations of Resyduo.

A novel tag-aware recommendation model named Tag Graph Convolutional Network (TGCN) has been proposed in \cite{chen_tgcn_2020} to reduce the ambiguity and redundancy issues. Built on top of a Collaborative Tag Graph (CTG), the approach combines neighbor sampling and aggregation strategies to extract informative features from heterogeneous neighbors belonging to the graph. In such a way, the semantics among the items is preserved and the obtained recommendations outperform existing approaches. 

Chen \etal \cite{chen_jit2r_2020} propose JIT2R, a hybrid framework that combines the traditional item-based prediction model with bootstrapping strategy. In the early stage, the proposed solution exploits the traditional softmax layer to predict the inial set of tags. Then, it iteratively labels item-tag pairs to improve the underpinning prediction model by using bootstrapping. The results show that JIT2R outperforms two state-of-the-art approaches in terms the computed accuracy metrics.

Although the process is similar, \RD is the first approach that exploits tags to recommend hardware components in the \iot domain.
%Rubei et al. in \cite{endowingTPL} propose an approach to improve the recommendations of TPLs by exploiting explicit user feedback. The approach is based on the Learning to Rank model, a supervised learning technique used to cope with the ranking task, and applied on \textsc{CrossRec} RS.The LTR model used is the Weighted Approximate-Rank Pairwise (WARP) which produces better performance on Top-N recommendations. In This approach, the user can express three types of feedback: positive, negative, or additive; the data are extracted and used to feed the model. 

\subsection{Recommender systems for TPLs}

Thung et al. in \cite{thung2013automated} propose LibRec, a tool that combines a CF-based algorithm and  association mining to retrieve popular Java libraries. First, the mining component suggests a set of libraries by relying on usage patterns. Then, the CF-based component suggests an additional set of TPLs that are used in similar projects.

LibSeek \cite{he2020diversified} provides diversified TPL for mobile applications. It exploits the matrix factorization technique to encode the relationships between apps and libraries. Furthermore, the approach is capable of handling the popularity bias by using an adaptive weighting scheme. 

%for using third-party libraries; this RS is based on a collaborative filtering approach. The approach is to mine information about software projects and find similarities between projects based on the libraries used. In LibRec the RS can suggest a set of libraries only if the project has already included at least one library. 

CrossRec~\cite{CrossRec_2021} exploits a graph-based structure that embodies mutual relationships between Java TPLs and corresponding Github projects. The system extracts matrixes that have been used to feed the underpinning CF-based engine. CrossRec eventually suggests a list of libraries ranked by a similarity score.

%It works based on a collaborative filtering engine to rank the outcomes. 
Req2Lib \cite{9054865} suggests relevant TPLs starting from the textual requirements extracted from StackOverflow posts. To overcome the cold-start problem of CF-based systems, the system employs a doc2vec pre-trained model and a sequence-to-sequence network to perform the recommendations.
% according to the cosine similarity.
Li \etal propose an approach based on graph neural networks (GNN), namely GRec, to recommend Android libraries \cite{10.1145/3468264.3468552}. As the first step, the system encodes both low and high-level app-library interactions in a graph-based structure. Grec eventually exploits the underpinning GNN to recommend diversified TPLs given the initial Android project.

Compared with the aforementioned systems, \RD is the first approach that recommends \AD software libraries by relying on predefined components. Furthermore, it handles the cold-start problem of the CF-based technique using the Type I and II recommendations capabilities to retrieve the needed components at the early stage of development.

\section{Conclusion}
\label{sec:conclusion}
To support the specification of \iot system, several MDE-based approaches have been devised. In this paper, we propose \RD, a recommender system that combines the MDE paradigm with traditional collaborative filtering algorithms to support \AD project development. First, we conceive a specific data model that represents all the peculiar metadata of a \AD project stored in a curated repository, \ie \PH. The specified model has been used to collect all the relevant data and feed the underpinning CF-based engine.

Furthermore, we apply the grid search to automatically set the optimal hyperparameters, including the algorithm type and the cutoff value. The system eventually provides there different types of recommendations that allow the users to configure their \AD project by selecting \iot hardware components and the corresponding software libraries.

Due to the lack of a proper baseline, we evaluate \RD by adopting the well-known ten-fold cross-validation and computing a set of accuracy and ranking metrics. Our findings show that \RD achieves better performances while \typeIII recommendations are considered, \ie software libraries, even though all the metrics values are far to be optimal. Nonetheless, the adopted methodology can be improved to increase the effectiveness of \RD, \eg by curating the training data by enhancing the proposed metamodel. Concerning the data gathering, we can \textit{i)} exploit the mined source code to suggest additional software artifacts, \eg API function call and \textit{ii)} mining and preprocessing additional data sources, \eg dedicated Q\&A post. Last but not least, we will improve the usability for non-expert user by embodying user feedback in the web interface.

%For its realization, existing approaches used in other domains, such as E-Commerce, Social Networks, and Software %Engineering, have been adapted to the specific Arduino's domain. For instance, the classic user-item approach %used in E-Commerce has been adapted for the Type I recommendations of components. The tag-aware recommendations %are mainly used in social networks, and finally, the third-party library recommendations are usually used in %RSSEs.
%Concerning the results obtained by ReSyDuo, in the Type I recommendations, the best result obtained is 27\% of %success rate. It is quite low value; this problem is due to the fact that users freely enter many components, and %this problem creates redundancies in the components' information.
%In the Type II recommendations, The best result obtained is a success rate of 39\% , an encouraging result. The best result is obtained for the recommendations of Type III. ReSyDuo is capable of recommending libraries with a success rate of 59\%; this result is very encouraging. ReSyDuo, in action, produces consistent recommendations in Type I and Type II despite the low success rate; this is because it can produce reliable recommendations for the most common components and tags.  For future work, it is crucial to increase the performance of ReSyDuo. For instance, a solution to the problem of information redundancy can be solved using Clustering techniques in data pre-processing. Furthermore, endowing ReSyDuo with an explicit user feedback mechanism can increase its performance.

\section*{Acknowledgment}
\addcontentsline{toc}{section}{Acknowledgment}

This work has been partially supported by the EMELIOT national research project, which has been funded by the MUR under the PRIN 2020 program (Contract 2020W3A5FY).
\cready{The authors would like to thank Andrea Serafini for his hard work in supporting the ResyDuo development and evaluation.}
\balance
\bibliographystyle{IEEEtran}  
\bibliography{main}

\end{document}